\documentclass{appolb}
\usepackage{epsfig}

\def\Jo#1#2#3#4{{#1} {\bf #2}, #3 (#4)}

\def\NPB{{Nucl. Phys.} {\bf B}}
\def\PLB{{Phys. Lett.}  {\bf B}}
\def\PRD{{Phys. Rev.} {\bf D}}
\def\PRP{{Phys. Rep.}}
\def\PRL{Phys. Rev. Lett.} 
\def\JHEP{J. High Energy Phys.} 
\def\IJMP{Int. J. Mod. Phys. {\bf A}} 

\def\ra{\rightarrow}
\def\be{\begin{equation}}
\def\ee{\end{equation}}
\def\gs{\mathrel{
   \rlap{\raise 0.511ex \hbox{$>$}}{\lower 0.511ex \hbox{$\sim$}}}}
\def\ls{\mathrel{
   \rlap{\raise 0.511ex \hbox{$<$}}{\lower 0.511ex \hbox{$\sim$}}}}
\newcommand{\obb}{0\mbox{$\nu\beta\beta$}}
\newcommand{\onbb}{neutrinoless double beta decay}

\newcommand{\dma}{\mbox{$\Delta m^2_A$}}
\newcommand{\dms}{\mbox{$\Delta m^2_\odot$}}
\newcommand{\ba}{\begin{array}{c}}  
\newcommand{\bad}{\begin{array}{ccc}}
\newcommand{\bea}{\begin{equation} \begin{array}{c}}
\newcommand{\eea}{ \end{array} \end{equation}}
\newcommand{\ea}{\end{array}}
\newcommand{\D}{\displaystyle} 
\newcommand{\meff}{\mbox{$\langle m \rangle$}}

\begin{document}

\pagestyle{plain}

\newcount\eLiNe\eLiNe=\inputlineno\advance\eLiNe by -1   
\title{
\vspace*{-2cm}\rightline{\rm hep-ph/0110258}
\rightline{DO-TH 01/13}\vspace{1.5cm} 
Leptogenesis, neutrinoless double beta decay 
and terrestrial $CP$ violation%
\thanks{Presented at XXV International School of Theoretical Physics, 
``Particle Physics and Astrophysics --- Standard Models and beyond'', 
Ustro\'n, Poland, 10-16 September 2001.}%
}
\author{Werner Rodejohann 
\address{Institut f\"ur Theoretische Physik, Universit\"at Dortmund,\\ 
Otto--Hahn--Str.\ 4, 44221 Dortmund, Germany\\
{\tt rodejoha@xena.physik.uni-dortmund.de}}}
\maketitle

\begin{abstract}
Leptogenesis in left--right symmetric theories is studied. The 
usual see--saw mechanism is modified by the presence of 
a left--handed Higgs triplet. A simple connection between the properties 
of the light left--handed and heavy right--handed neutrinos 
is found. Predictions of this scenario for neutrinoless double 
beta decay and terrestrial $CP$ violation in long--baseline 
experiments are given. These observables can in principle distinguish 
different realizations of the model. 
\end{abstract}


\section{\label{sec:intro}Introduction}
One of the problems waiting to 
be solved in particle physics and cosmology 
is the explanation of the baryon asymmetry of the universe. 
Since Standard Model baryogenesis fails to produce a 
sufficient baryon asymmetry, 
other, new physics approaches are being followed. 
Among them towers out leptogenesis \cite{leptogenesis} as 
one of the most popular. 
Heavy right--handed Majorana neutrinos violate 
$CP$ and lepton number during their out--of--equilibrium decay, 
thereby  --- when sphalerons \cite{sphaleron} 
convert the lepton asymmetry in a baryon asymmetry --- 
fulfilling all of Sakharov's three conditions \cite{sakharov}.\\ 
The impressive evidence for non--vanishing neutrino masses 
opens now the possibility to study this new physics problem on 
a broader phenomenological basis. Typical models build to explain the 
neutrino mass and mixing scheme predict also heavy right--handed Majorana 
neutrinos, mostly due to some see--saw \cite{seesaw} mechanism. 
It is now a fruitful question to ask if a given model for neutrino masses 
also explains the baryon asymmetry via the leptogenesis mechanism. 
A number of groups have studied this within their respective approach 
\cite{others}.\\ 
As the name already indicates, left--right (LR) symmetric theories 
represent a natural way to connect the light left--handed with 
the heavy right--handed neutrino sector. In \cite{JPR1} the 
relationship of both sectors and the impact on leptogenesis 
was analyzed. An observable effect 
of the relation between neutrino oscillation and leptogenesis 
was then proposed in \cite{JPR2}. The three yet unknown phases in the 
left--handed neutrino mass matrix govern the magnitude of the 
effective neutrino mass measured in neutrinoless double beta decay and 
the size of terrestrial 
$CP$ violating effects in long--baseline experiments. 
Many models explain the baryon 
asymmetry as well as the light mass and mixing scheme. Predictions 
of other observables are then very helpful to rule out 
or confirm models. The relationship of terrestrial $CP$ violation and 
leptogenesis was also analyzed in \cite{othersCP}.\\
The paper is organized as follows: In Section \ref{sec:form} the 
connection of leptogenesis and neutrino oscillation 
in left--right symmetric theories 
is given and the results on the baryon asymmetry are presented. 
The connection to terrestrial $CP$ violation is made in 
Section \ref{sec:CP} and the conclusions are drawn 
in Section \ref{sec:concl}.

\section{\label{sec:form}Neutrino oscillation and leptogenesis in 
left--right symmetric theories}
\begin{figure}[ht]
\begin{center}
\hspace{-1.cm}
\epsfig{file=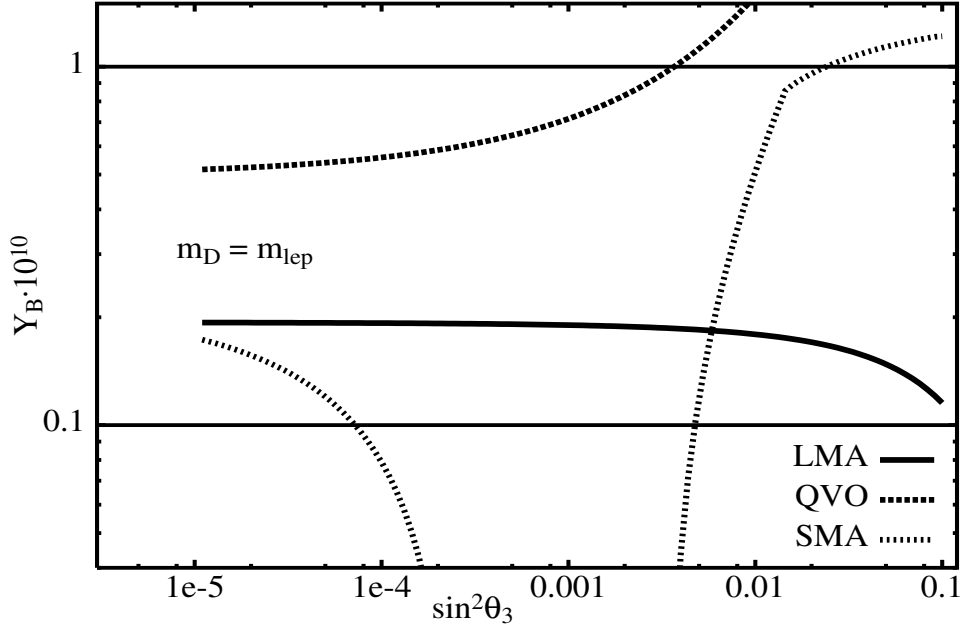,width=13cm,height=8.5cm} 
\caption{\label{fig:mlep}Baryon asymmetry as a function of $s_3$ 
for $\tilde{m}_D = m_{\rm lep}$ and all three solar solutions. 
We chose $3 \alpha = 4 \beta = 6 \delta = \pi$, 
$\dma = 3.2 \cdot 10^{-3}$ eV$^2$ and $\tan_2^2 = 1$. For the 
solar solutions, we took $\dms = 5 \cdot 10^{-6}$ eV$^2$ and 
$\tan_1^2 = 5 \cdot 10^{-4}$ for SMA,  $\dms = 5 \cdot 10^{-5}$ eV$^2$ and 
$\tan_1^2 = 1$ for LMA and $\dms = 10^{-8}$ eV$^2$ and 
$\tan_1^2 = 1$ for QVO\@. The smallest mass state is $m_1=10^{-5}$ eV for QVO 
and $m _1 = 10^{-4}$ eV for SMA as well as LMA\@.}
\end{center}
\end{figure}

In LR symmetric theories the see--saw formula reads 
\be \label{eq:mnu}
m_\nu = m_L - \tilde{m}_D \, M_R^{-1} \, \tilde{m}_D^T \, , 
\ee 
where $m_L$ and $M_R$ are Majorana mass matrices generated by Higgs 
triplets and $\tilde{m}_D$ is a Dirac mass matrix. 
The matrix $m_\nu$ is further diagonalized by
\be \label{eq:uldef}
U_L^T \, m_\nu \, U_L = {\rm diag} (m_1,m_2,m_3) ,
\ee
where $m_i$ are the light neutrino masses. 
The symmetric matrix $M_R$ also appears in the Lagrangian
\be \label{eq:convss}
-\mbox{$\cal{L}$}_Y
= \overline{l_{iL}} \, \frac{\D \Phi}{\D v} \,
\tilde{m}_{D ij} \, N'_{Rj}
+ \frac{\D 1}{\D 2}\overline{N'^c_{Ri}} \, M_{R ij} \, N'_{Rj} + \; \rm h.c.
\ee
with $l_{iL}$ the leptonic doublet and $v \simeq 174$ GeV the
vacuum expectation value (vev) of the Higgs doublet $\Phi$.
Diagonalizing $M_R$ brings us to the physical basis           
\be \label{eq:urdef}
U_R^T \, M_R \, U_R = {\rm diag}(M_1, M_2, M_3) \, . 
\ee
The asymmetry is caused by the interference of tree level with one--loop
corrections for the decays of the lightest Majorana, $N_1 \ra \Phi \, l^c$
and $N_1 \ra \Phi^\dagger \, l$:
\bea \label{eq:eps}  
\varepsilon = \frac{\D \Gamma (N_1 \ra \Phi \, l^c) -
\Gamma (N_1 \ra \Phi^\dagger \, l)}{\D \Gamma (N_1 \ra \Phi \, l^c) +
\Gamma (N_1 \ra \Phi^\dagger \, l)} \\[0.3cm]
= \frac{\D 1}{\D 8 \, \pi \, v^2} \frac{\D 1}{\D (m_D^\dagger m_D)_{11}}
\sum\limits_{j=2,3} {\rm Im} (m_D^\dagger m_D)^2_{1j} \, f(M_j^2/M_1^2) \, . 
\eea
The function $f$ includes
terms from vertex and self--energy contributions:
\be \label{eq:fapprox}
f(x) = \sqrt{x} \left(1 + \frac{1}{1 - x} -
(1 + x) \, \ln \left(\frac{1 + x}{x}\right) \right)
\simeq - \frac{3}{2 \, \sqrt{x}}  .
\ee
The approximation holds for $x \gg 1$.\\
In our approach, the left--right symmetry \cite{LR} plays an important role.
It relates the unitary matrices $U_L$ and $U_R$ to each other since the 
triplet induced Majorana mass matrices in Eq.\ (\ref{eq:mnu}) 
have the same coupling matrix $f$ in generation space: 
\be \label{eq:mlmr}
\bad
m_L = f \, v_L & \mbox{ and } & M_R = f \, v_R   \; .
\ea
\ee
The numbers $v_{L, R}$ are the vevs of the left-- and right--handed
Higgs triplets, whose existence is needed to maintain the left--right
symmetry. They receive their vevs at the minimum of the potential,
producing at the same time masses for the gauge bosons.
In general \cite{LR}, this results in
\be \label{eq:vlvr}
v_L \, v_R \simeq \gamma \, v^2 ,
\ee
where the constant $\gamma$ is a model dependent parameter of
$\cal{O}$(1). Inserting this equation as well as Eq.\ (\ref{eq:mlmr}) in
(\ref{eq:mnu}) yields
\be \label{eq:mnulr}
m_\nu = v_L \,
\left( f - \tilde{m}_D \,
\frac{f^{-1}}{\gamma \, v^2} \, \tilde{m}_D^T \right) .
\ee                                                  
\\
If one 
compares the relative magnitude of the two contributions
in Eq.\ (\ref{eq:mnu}), 
denoting the largest mass in the Dirac matrix with $m$,
one finds that 
\be \label{eq:estlr}
\frac{\D |\tilde{m}_D \, M_R^{-1} \, \tilde{m}_D^T |}{\D |m_L| } \simeq
\frac{m^2/v_R}{\D v_L} \simeq
\frac{\D m^2}{\D \gamma \, v^2}\, . 
\ee 
Here, we only used Eq.\ (\ref{eq:vlvr}) and assumed that
the matrix elements of $f$ and $f^{-1}$ are of the same order of
magnitude. It is seen that this ratio
is of order one only for the top quark mass, \ie if one identifies
the Dirac mass matrix with the up quark mass matrix.\\
We finally specify the order of
magnitude of $v_{L,R}$. The scale of
$m_\nu=v_L \, f$ is $10^{-2} \ldots 10^{-3}$ eV, which --- for not too 
small $f$ --- is
only compatible with  $v_L\, v_R \simeq \gamma \, v^2$ 
for $v_R \simeq 10^{14} \ldots 10^{15}$ GeV\@.
This means that $v_R$ is close to the 
grand unification scale and $v_L$ is of the order of the neutrino masses, 
which is expected since $m_L$ is the dominating contribution to
$m_\nu$. In the following, $v_R = 10^{15}$ GeV and $\gamma=1$ is 
assumed.\\
From the decay asymmetry $\varepsilon$ the baryon asymmetry $Y_B$ is 
obtained by 
\be \label{eq:YBth}
Y_B = c\, \kappa \frac{\varepsilon}{g^\ast} ,
\ee
where $c \simeq -0.55$ is the fraction of the lepton asymmetry 
converted to a baryon asymmetry via sphaleron 
processes \cite{hartur}, 
$\kappa$ a suppression factor due to lepton--number 
violating wash--out processes (see \cite{hiki} for an improved fit) 
and $g^\ast \simeq 110$ 
the number of massless degrees of freedom at the time of the decay.
Experimentally, the preferred range for the asymmetry is \cite{citeYB} 
$ Y_B \simeq (0.1 \ldots 1) \cdot 10^{-10} $.\\ 
The strategy goes as follows: 
In Eq.\ (\ref{eq:mnulr}) one inserts the solar solution, \ie the 
small angle (SMA), large angle (LMA) or quasi--vacuum (QVO) 
solution, see \eg \cite{carlos}. 
The light neutrino masses $m_i$ are obtained by assuming the   
hierarchical scheme. The Dirac mass matrix $\tilde{m}_D$ 
can be expected to be an up (down) quark or lepton mass matrix, denoted 
$m_{\rm up}$, $m_{\rm down}$ and $m_{\rm lep}$, respectively. 
Eq.\ (\ref{eq:mnulr}) is then solved for $f = M_R/v_R$ and 
$M_R$ is diagonalized to obtain the baryon asymmetry via 
Eqs.\ (\ref{eq:eps},\ref{eq:YBth}).\\
Performing a random scan 
of the allowed oscillation parameters and 
the three phases, 
it is found that if $\tilde{m}_D$ is a down quark or 
lepton mass matrix, $m_1$ should not be too small, 
\ie larger than $10^{-6} $ eV\@. 
The LMA 
solution gives in more cases the correct baryon asymmetry and is thus  
slightly favored over SMA and QVO\@. 
If $\tilde{m}_D$ is an up quark mass matrix, 
fine tuning of the parameters is required. Due to the 
large hierarchy of the quark and lepton masses, 
it is sufficient to use a mass matrix which has 
just the heaviest mass as the (33) entry. Fig.\ \ref{fig:mlep} 
shows $Y_B$ in case of $\tilde{m}_D = m_{\rm lep}$.\\  
If we identify $\tilde{m}_D$ with the down quark or charged lepton mass
matrix, then the ratio in Eq.\ (\ref{eq:estlr}) is always much smaller than
one, so that the second term in Eq.\ (\ref{eq:mnulr}) can be 
neglected and it follows \cite{JPR2}
\be \label{eq:fsimnu}
f \simeq \frac{1}{v_L} \, m_\nu .
\ee
Therefore, with the help of 
Eqs.\ (\ref{eq:uldef},\ref{eq:urdef},\ref{eq:mlmr}), one arrives at
a very simple connection between the left-- and right--handed
neutrino sectors: 
\be \label{eq:cons}
U_R = U_L \mbox{ and } M_i = m_i \, \frac{v_R}{v_L} \, . 
\ee
The striking property is that the light neutrino masses are
proportional to the heavy ones. Analytical estimates for the 
baryon asymmetry can now be 
performed. We work with a convenient parametrisation of $U_L$, 
\bea \label{eq:Upara}
U_L = U_{\rm CKM} \cdot P = U_{\rm CKM} \;
{\rm diag}(1, e^{i \alpha}, e^{i (\beta + \delta)}) \\[0.3cm]
= \left( \bad
c_1 c_3 & s_1 c_3 & s_3 e^{-i \delta} \\[0.2cm]
-s_1 c_2 - c_1 s_2 s_3 e^{i \delta}
& c_1 c_2 - s_1 s_2 s_3 e^{i \delta}
& s_2 c_3 \\[0.2cm]
s_1 s_2 - c_1 c_2 s_3 e^{i \delta} &
- c_1 s_2 - s_1 c_2 s_3 e^{i \delta}
& c_2 c_3\\
               \ea   \right)
 {\rm diag}(1, e^{i \alpha}, e^{i (\beta + \delta)}) ,
\eea
where $c_i = \cos\theta_i$, $s_i = \sin\theta_i$ and the diagonal matrix 
$P$ contains the additional two Majorana phases $\alpha$ and $\beta$.  
Assuming maximal atmospheric and solar mixing, 
$c_1^2 = c_2^2 = 1/2$, and taking only the leading order in $s_3$, 
one finds for the LMA and QVO solutions \cite{JPR2} 
\be \label{eq:estYBLM}
\bad Y_B \cdot 10^{10} \ls 4.1 \frac{\D 1}{\D 1 - 2\, s_3 \, c_\delta}
\left( \frac{\D m}{\D \rm GeV} \right)^2 
\left\{
(s_{2\alpha} + 4 \, s_3 \, s_{\delta} \, c_{2\alpha}   )
\frac{\D m_1}{\D \sqrt{\dms}} \right. \\[0.5cm] 
\left. + 2 (s_{2(\beta + \delta) }- 2 \, s_3\, s_{2\beta+\delta}) 
\frac{\D m_1}{\D \sqrt{\dma}} 
\right\}  \, ,  
\ea  
\ee
where $c_\delta = \cos \delta$, 
$s_{2 \alpha} = \sin 2 \alpha$ and so on. The solar (atmospheric) 
$\Delta m^2$ is denoted \dms{} (\dma). 
It is seen 
explicitly that $Y_B$ 
vanishes if $CP$ conservation holds, \ie if all phases are zero or $\pi$. 
The asymmetry is proportional 
to the square of the heaviest entry in $\tilde{m}_D$, \ie 
the tau or bottom quark mass. 
Furthermore, $Y_B$ is 
proportional to the lightest neutrino mass eigenstate $m_1$, which can 
be used to set a lower limit on it, it is of the order $10^{-7}$ 
to $10^{-8}$ eV\@.\\
If  $\tilde{m}_D = m_{\rm up}$ then $m_\nu$ receives a contribution 
from the conventional see--saw term $\tilde{m}_D M_R^{-1} \tilde{m}_D^T$ 
and the proportionality on $m_1$ vanishes, see \cite{JPR1} for 
details.

\section{\label{sec:CP}Terrestrial $CP$ violation}
The remaining unknowns in this approach 
are the three $CP$ violating phases in the mixing matrix $U_L$ and 
the size of the smallest mass eigenstate $m_1$. Within the 
parametrisation Eq.\ (\ref{eq:Upara}) the phases $\alpha$ and $\beta$ 
govern the magnitude of \onbb. 
The third phase $\delta$ is responsible for $CP$ violating effects in 
oscillation experiments.\\
The latest SuperKamiokande \cite{SK} and first SNO \cite{SNO} 
data favor LMA over the other solar solutions. This is good news since 
leptonic $CP$ violation in long--baseline experiments can 
only be measured if nature has chosen LMA\@. 
Effects of $CP$ violation are proportional to the rephasing invariant 
determinant $J_{CP}$ \cite{jarlskog}, which shows up \eg in the 
difference of the $CP$ conjugated oscillation 
probabilities 
\bea \label{eq:JCP}
P(\nu_e \ra \nu_\mu) - P(\bar{\nu}_e \ra \bar{\nu}_\mu) \propto J_{CP} = 
\frac{1}{8} \, \sin 2 \theta_1 \, \sin 2 \theta_2 
\, \sin 2 \theta_3 \, \cos \theta_3 \, \sin \delta \\[0.3cm]
\le \frac{1}{4} \, \sin \theta_3 (1 - \sin^2 \theta_3 ) \, . 
\eea
In addition, the higher \dms{} is, 
the higher are the prospects for detecting 
the $CP$ violation \cite{martin}, though the 
details depend on the experimental facilities.\\
\begin{figure}[b]
\begin{center}
\hspace{-4cm}
\epsfig{file=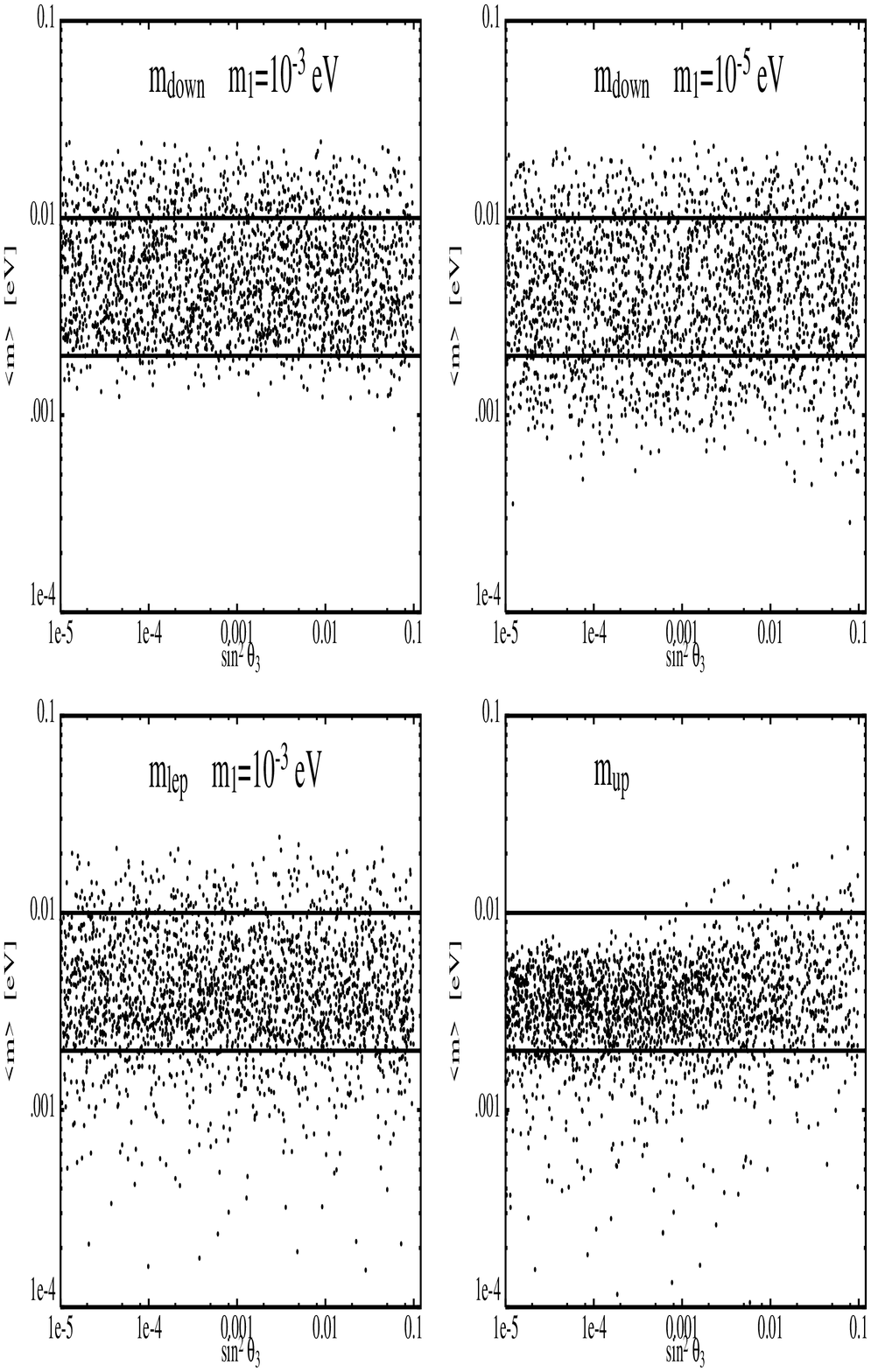,width=15cm,height=8.5cm} 
\caption{\label{fig:meff}Distribution of events 
in the \meff--$\sin^2 \theta_3$ plane for the LMA solution, 
different $m_1$ and $\tilde{m}_D$. }
\end{center}
\end{figure}
In the hierarchical mass scheme, 
LMA also provides the highest Majorana mass for 
the electron neutrino, which can be measured through 
\onbb{} (\obb). It is defined as 
\be \label{eq:meff}
\meff = \sum_i \, U_{L ei}^2 \, m_i
\ee   
and due to the complex matrix elements $U_{L \alpha i}$ there is the 
possibility of cancellation \cite{ichNPB} 
of terms in Eq.\ (\ref{eq:meff}).\\ 
The quantities \meff{} and $J_{CP}$ are observables, which are 
depending on the $CP$ violating phases which also govern 
the lepton asymmetry. 
It is therefore interesting to ask if the parameters
that produce a satisfying $Y_B$ also deliver sizable 
\meff{} and/or $J_{CP}$. 
To study this, a random scan of the allowed variables of the LMA 
solution was performed. 
The highest fraction of parameter sets providing sufficient 
$Y_B$ occurs for high $m_1$ and a ``low'' Dirac mass matrix, \ie 
$\tilde{m}_D$ should be a lepton (43 $\%$) or down quark 
(23 $\%$) mass matrix. It is interesting to note that in the most 
simple realization of LR models $\tilde{m}_D$ 
is the charged lepton mass matrix. 
For lower $m_1$ or $\tilde{m}_D = m_{\rm up}$ 
the fraction of parameters producing a correct asymmetry 
decreases to less than 5 $\%$. As mentioned, basically no $m_1$ dependence 
exists for $\tilde{m}_D = m_{\rm up}$.  
Approximately all the 
parameter sets providing a correct asymmetry also produce \meff{} 
bigger than $2 \cdot 10^{-3}$ eV, the lowest limit achievable by 
the GENIUS project \cite{GENIUS}. For $m_1 = 10^{-3}$ eV, 
about 4 $\%$ of the parameter sets give \meff{} bigger than 
0.01 eV\@. Fig.\ \ref{fig:meff} shows the distribution of events in the 
\meff--$\sin^2 \theta_3$ plane. The difference for different 
cases is easily seen.\\
\begin{figure}[t]
\begin{center}
\hspace{-4cm}
\epsfig{file=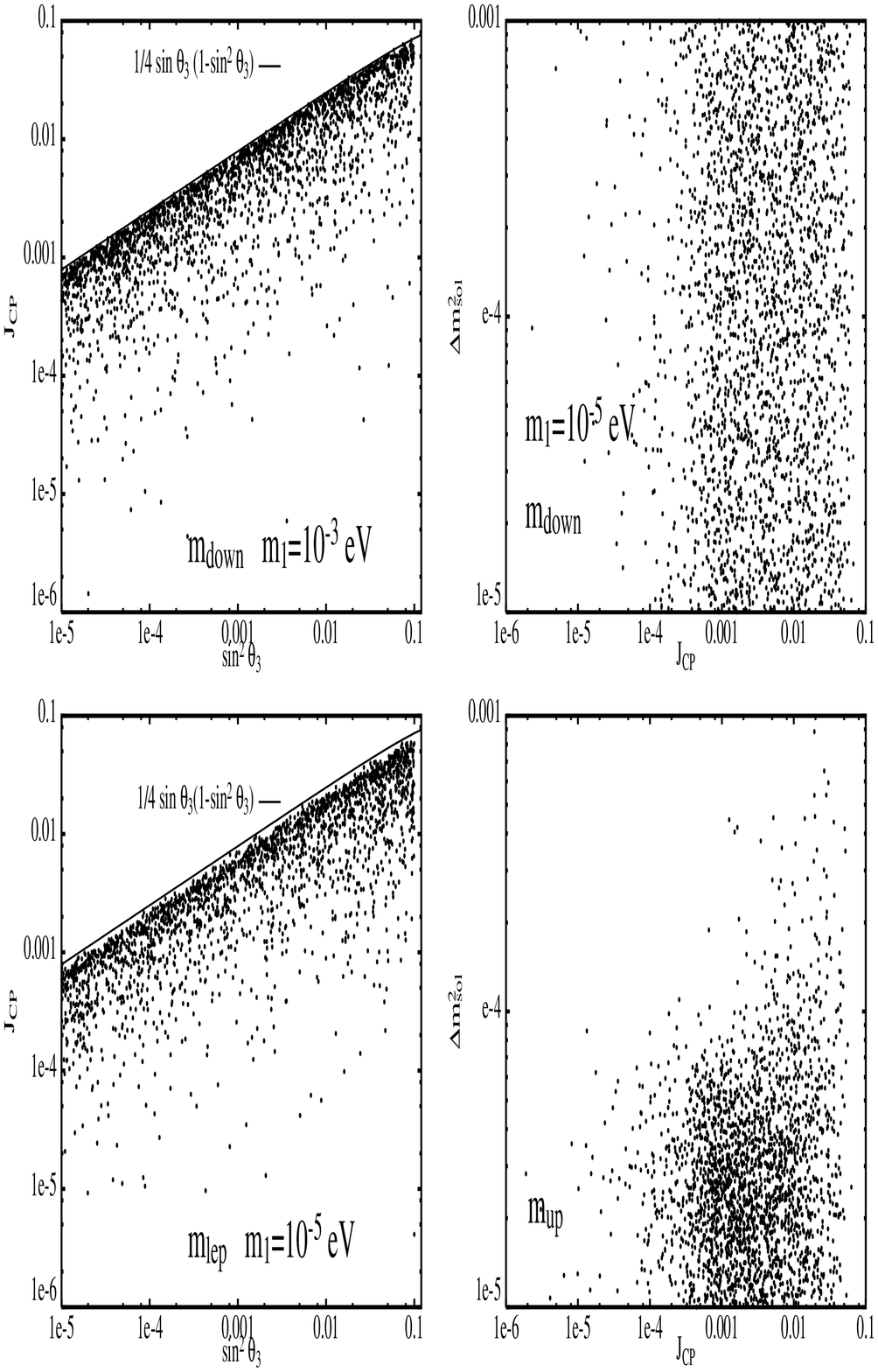,width=15cm,height=8.5cm} 
\end{center}
\vspace{-1cm}
\caption{\label{fig:JCP}Distribution of events in the 
$J_{CP}$--$\sin^2 \theta_3$ (left) and $J_{CP}$--\dms{} (right) 
plane for the LMA solution, different $m_1$ different $\tilde{m}_D$. }
\end{figure} 
Regarding $CP$ violation, a 
criterion for observability might be $J_{CP} \ge 10^{-4}$  
and $\dms{} \ge 10^{-4}$ eV$^2$. Approximately half of the 
events that give sufficient $Y_B$ also fulfill these constraints. 
Therefore, again high $m_1$ and $\tilde{m}_D = m_{\rm down}$ 
or $m_{\rm lep}$ are required to expect measurable $CP$ violation. 
Fig.\ \ref{fig:JCP} shows the distribution of 
$J_{CP}$ against $\sin^2 \theta_3$ and \dms, respectively. 
Again, the difference is easily seen. The case 
$\tilde{m}_D = m_{\rm up}$ favors low \dms. 

\section{\label{sec:concl}Conclusions}
Leptogenesis in left--right symmetric models is studied. 
A simple formula for $Y_B$ can be derived, 
expressing the baryon asymmetry in terms of oscillation parameters 
and $CP$ violating phases. In many cases a sufficient baryon asymmetry 
is produced and the LMA solution is favored.  
Many models in this scenario as well as other 
approaches fulfill these constraints. 
In search for an additional criterion we 
therefore apply our model also to \obb{} and 
terrestrial $CP$ violating effects in long--baseline experiments. 
In order to expect a sizable signal in \obb{} and measurable $CP$ 
violating effects in long--baseline experiments, $m_1$ of order 
$10^{-3}$ eV is required, and $\tilde{m}_D$ should be 
a lepton or perhaps a down quark mass matrix. The low energy 
observables $J_{CP}$ and \meff{} can in principle be used to distinguish 
these possibilities and could also be used to distinguish 
other leptogenesis models.\\
Baryon number 
and $CP$ violation are necessary conditions for the generation of 
a baryon asymmetry. Since $Y_B$ gets converted from a lepton asymmetry, 
lepton number violation is required. Thus, 
\obb{} and terrestrial $CP$ violation 
provide a possibility to check two of Sakharov's conditions at 
low energy. Furthermore, given that in many models the heavy 
right--handed neutrinos may not be observable at realistic 
collider energies, \obb{} and terrestrial $CP$ violation 
could be useful to validate leptogenesis. 

\begin{center}
{\bf \large Acknowledgments}
\end{center}
I warmly thank my collaborators A.\ S.\ Joshipura and E.\ A.\ Paschos and 
am grateful to W.\ Buchm\"uller and K.\ R.\ S.\ Balaji 
for helpful discussions. I would further 
like to thank the organizers of the XXV International School 
of Theoretical Physics for their hospitality.  
This work has been supported in part by the
``Bundesministerium f\"ur Bildung, Wissenschaft, Forschung und
Technologie'', Bonn under contract No. 05HT1PEA9. 
Financial support from the Graduate College
``Erzeugung und Zerf$\ddot{\rm a}$lle von Elementarteilchen''
at Dortmund university is gratefully acknowledged.

\end{document}